\documentclass[11pt]{article}
\usepackage{amssymb,latexsym,amsmath,amsbsy}
\usepackage{cite}
\usepackage{stmaryrd}  
\usepackage{arydshln}  
\usepackage{mathdots}  
\newtheorem{theo}{Theorem}

\newtheorem{prop}[theo]{Proposition}
\def\deg{\mathop{\rm deg}\nolimits}
\newcommand{\myatop}[2]{\genfrac{}{}{0pt}{}{#1}{#2}}
\def\mybox{\hfill$\Box$}
\newcommand{\overbar}[1]{\mkern 3.0mu\overline{\mkern-3.0mu#1\mkern-3.0mu}\mkern 3.0mu}

\renewcommand{\theequation}{\arabic{section}.{\arabic{equation}}}

\begin{document}
\begin{center}
{\Large \bf
Gel'fand-Zetlin basis for a class of representations\\[2mm] 
of the Lie superalgebra $\mathfrak{gl}(\infty|\infty)$ }\\[5mm] 
{\bf N.I.~Stoilova}\footnote{E-mail: stoilova@inrne.bas.bg}\\[1mm] 
Institute for Nuclear Research and Nuclear Energy,\\ 
Boul.\ Tsarigradsko Chaussee 72, 1784 Sofia, Bulgaria\\[2mm] 
{\bf J.\ Van der Jeugt}\footnote{E-mail: Joris.VanderJeugt@UGent.be}\\[1mm]
Department of Applied Mathematics, Computer Science and Statistics, Ghent University,\\
Krijgslaan 281-S9, B-9000 Gent, Belgium.
\end{center}

\addtolength{\parskip}{2mm}

\begin{abstract}
A new, so called odd Gel'fand-Zetlin basis is introduced for the irreducible covariant tensor representations of
the Lie superalgebra $\mathfrak{gl}(n|n)$. 
The related Gel'fand-Zetlin patterns are based upon the decomposition according to a particular chain of subalgebras of $\mathfrak{gl}(n|n)$.
This chain contains only genuine Lie superalgebras of type $\mathfrak{gl}(k|l)$ with $k$ and $l$ nonzero (apart from the final element of the chain which is $\mathfrak{gl}(1|0)\equiv \mathfrak{gl}(1)$).
Explicit expressions for a set of generators of the algebra on this Gel'fand-Zetlin basis are determined.
The results are extended to an explicit construction of
a class of irreducible highest weight modules of the general linear Lie superalgebra $\mathfrak{gl}(\infty|\infty)$. 
\end{abstract}

\vskip 10mm
\noindent Running title: GZ-basis for representations of  $\mathfrak{gl}(\infty|\infty)$ 



\setcounter{equation}{0}
\section{Introduction} \label{sec:Introduction}%

Many decades ago, Green~\cite{Green} generalized standard quantum statistics describing bosons and fer\-mions, and introduced parabosons and parafermions.
Typically, the relations for paraboson and para\-fer\-mion creation and annihilation operators are triple relations rather than just commutators or anticommutators.
Greenberg and Messiah~\cite{Greenberg} studied mixed system consisting of parafermions $f_j^\pm$ and parabosons $b_j^\pm$, and their relative commutation relations.
As far as the underlying algebraic structure is concerned, Palev~\cite{Palev1} proved that $k$ parafermions $f_j^\pm$ and $n$ parabosons $b_j^\pm$ with so-called relative parafermion relations generate the orthosymplectic Lie superalgebra $\mathfrak{osp}(2k+1|2n)$. 
The explicit construction of the corresponding parastatistics Fock space, which is an infinite-dimensional unitary representation of $\mathfrak{osp}(2k+1|2n)$, turned out to be a very difficult problem, and was completed only recently~\cite{SJ15}.
In this construction, the branching $\mathfrak{osp}(2k+1|2n)\supset \mathfrak{gl}(k|n)$ plays a crucial role, and allows to label the states of the parastatistics Fock space by means of Gel'fand-Zetlin (GZ) labels of covariant representations of the subalgebra $\mathfrak{gl}(k|n)$.

The real interest lies in such quantum systems with infinite degrees of freedom, consisting of a mixed system of an infinite number of parafermions and parabosons ($k\rightarrow\infty$ and $n\rightarrow\infty$).
In such a case, the parastatistics Fock space could be seen as a unitary representation of some infinite rank Lie superalgebra $\mathfrak{osp}(\infty|\infty)$. 
In order to study this representation along the lines of~\cite{parafermion,paraboson,SJ15}, it is clear that the subalgebra $\mathfrak{gl}(\infty|\infty)$ and its covariant representations will play a crucial role.
Of course, one has to be more precise in what is meant by the infinite rank Lie superalgebra $\mathfrak{gl}(\infty|\infty)$, and by its class of covariant representations.

For an infinite number of parabosons (or an infinite number of parafermions, but not mixed), the problem was already investigated in~\cite{SV2009}. 
Roughly speaking, for that case the GZ-basis for $\mathfrak{gl}(n)$ covariant representations could be naturally extended to ``a stable GZ-pattern'' for $\mathfrak{gl}(\infty)$, and that offered a way of describing the paraboson (resp.\ parafermion) Fock space with an infinite degree of freedom.
Turning our attention to a mixed system, one might expect that the known GZ-basis~\cite{SV2010} for covariant representations of $\mathfrak{gl}(k|n)$ can also quite naturally be extended to some infinite but stable GZ-patterns for $\mathfrak{gl}(\infty|\infty)$.
However, there is a problem with this.
Indeed, the natural GZ-basis for $\mathfrak{gl}(k|n)$~\cite{SV2010} is based on the decomposition according to the branching
\begin{equation}
\mathfrak{gl}(k|n) \supset \mathfrak{gl}(k|n-1) \supset \mathfrak{gl}(k|n-2) \supset \cdots \supset \mathfrak{gl}(k|1) \supset \mathfrak{gl}(k) \supset \mathfrak{gl}(k-1) \supset\cdots\supset \mathfrak{gl}(2) \supset \mathfrak{gl}(1).
\label{chain-mn}
\end{equation}
The corresponding GZ-basis pattern~\cite[eq.~(3.12)]{SV2010} consists of a triangular array of $(k+n)$ rows (and columns), in such a way that the triangle consisting of the $k$ bottom rows corresponds to a common $\mathfrak{gl}(k)$ GZ-pattern, the triangle corresponding to the $n$ rightmost columns corresponds to a $\mathfrak{gl}(n)$ GZ-pattern, and the remaining $(k\times n)$ rectangle contains the remaining basis labels satisfying certain ``$\theta$-conditions''.
Clearly, such a basis can not be used in the limit when both $k$ and $n$ tend to infinity.
For example, when $k$ goes to infinity, the bottom triangle (with, evidently, some stability conditions) stretches to infinity and one can never get control over the labels related to the $\mathfrak{gl}(n)$ part of $\mathfrak{gl}(k|n)$.
Similarly, when $n$ goes to infinity, the upper right triangle stretches to infinity at the left, and one cannot get hold of the labels related to the $\mathfrak{gl}(k)$ part of $\mathfrak{gl}(k|n)$.
This problem, described schematically only, is related to the particular choice of subalgebras in the chain~\eqref{chain-mn}.

In this paper, we show that another chain of subalgebras,
\begin{equation}
\mathfrak{gl}(n|n) \supset \mathfrak{gl}(n|n-1) \supset \mathfrak{gl}(n-1|n-1) \supset \mathfrak{gl}(n-1|n-2) \supset \mathfrak{gl}(n-2|n-2) \supset 
\cdots \supset \mathfrak{gl}(1|1) \supset \mathfrak{gl}(1) 
\label{chain-nn}
\end{equation}
is appropriate for considering the limit when $n$ goes to infinity, and allows us to construct stable GZ-patterns for covariant representations of $\mathfrak{gl}(\infty|\infty)$.
This is also related to the fact that the last chain can be ``reversed'',
\begin{equation}
\mathfrak{gl}(1)=\mathfrak{gl}(1|0) \subset \mathfrak{gl}(1|1) \subset \mathfrak{gl}(2|1) \subset \mathfrak{gl}(2|2) 
\subset \mathfrak{gl}(3|2) \subset \mathfrak{gl}(3|3) \subset \cdots \subset \mathfrak{gl}(\infty|\infty).
\label{chain-nni}
\end{equation}

The main part of the paper is thus devoted to constructing a new GZ-basis for covariant representations of $\mathfrak{gl}(n|n)$ according to the chain of subalgebras~\eqref{chain-nn}.
Of course, the notion of ``GZ-basis'' does not just refer to a way of labeling the basis vectors of the representation, but also to the fact that the action of a set of generators of the algebra on these basis vectors can be given.
In the current situation, it turns out that the set of generators for which the action on the new GZ-basis vectors takes the simplest form, is a set of non-distinguished simple root vectors of $\mathfrak{gl}(n|n)$ consisting of odd roots only.
For this reason, the new GZ-basis will be called ``the odd GZ-basis''.
Our main result is Theorem~\ref{theo4}, giving the explicit action of the generators on the odd GZ-basis vectors.

The new GZ-basis for $\mathfrak{gl}(n|n)$ is interesting on its own: it is simple, elegant, and for the first time one is dealing with a basis on which the action of a set of odd generators is simple and natural.
The main importance is that the basis and generator action can be extended to a class of representations of $\mathfrak{gl}(\infty|\infty)$.

There are a number of ways to define the infinite dimensional and infinite rank Lie superalgebra $\mathfrak{gl}(\infty|\infty)$.
We follow the common definition~\cite{Kac3} in terms of certain squared infinite matrices with a finite number of nonzero entries, on which a grading is fixed.
The GZ-patterns for $\mathfrak{gl}(n|n)$ can be extended to infinite GZ-patterns for $\mathfrak{gl}(\infty|\infty)$ in a rather straight way.
Those infinite GZ-patterns should however satisfy a certain stability condition, in order to correspond to a basis for an irreducible representation of $\mathfrak{gl}(\infty|\infty)$.

In Section~\ref{sec:glmn} we define the Lie superalgebra $\mathfrak{gl}(k|n)$ and remind the reader of covariant tensor 
representations of $\mathfrak{gl}(k|n)$. In Section~\ref{sec:representations}, we construct the covariant tensor representations of $\mathfrak{gl}(n|n)$
introducing the odd Gel'fand-Zetlin basis following the decomposition~(\ref{chain-nn}). We give the action of the $\mathfrak{gl}(n|n)$ generators
on the basis and prove that the algebra relations are satisfied in these representations. The infinite-dimensional general linear Lie 
superalgebra  $\mathfrak{gl}(\infty|\infty)$ is defined in Section~\ref{sec:gl(InftyInfty)}. Infinite odd GZ-patterns with a so-called stability
index are introduced in Section~\ref{sec:infty representations}. We prove that the Lie superalgebra  $\mathfrak{gl}(\infty|\infty)$ has an 
irreducible action on basis vectors labelled by these stable odd GZ-patterns.

\setcounter{equation}{0}
\section{The Lie superalgebra $\mathfrak{gl}(k|n)$ and covariant representations} \label{sec:glmn} 

The underlying vector space for the Lie superalgebra ${\mathfrak g}=\mathfrak{gl}(k|n)$ consists of the space of $(r\times r)$-matrices, with
\begin{equation}
r=k+n.
\end{equation}
The Lie superalgebra ${\mathfrak g}=\mathfrak{gl}(k|n)$ can be defined~\cite{Kac1,Kac2} through its natural matrix realization
\begin{equation}
\mathfrak{gl}(k|n)=\{ x=\left(\begin{array}{cc} A&B\\ C&D\end{array}\right)
| A\in M_{k\times k}, B\in M_{k\times n}, C\in M_{n\times k},
  D\in M_{n\times n} \},
\label{defgl}
\end{equation}
where $M_{p\times q}$ is the space of all $p\times q$ complex matrices.
The even subalgebra $\mathfrak{gl}(k|n)_{\bar 0}$ has $B=0$ and $C=0$; the odd
subspace $\mathfrak{gl}(k|n)_{\bar 1}$ has $A=0$ and $D=0$. 
Note that $\mathfrak{gl}(k|n)_{\bar 0} = \mathfrak{gl}(k)\oplus \mathfrak{gl}(n)$. 
The Lie superalgebra is then defined by means of the bracket $\llbracket x,y \rrbracket=xy-(-1)^{\deg(x)\deg(y)}yx$, where $x$ and $y$ are homogeneous elements.

It will be convenient to use the ordered set $\{-k,\ldots,-2,-1;1,2,\ldots n\}$ as indices for the rows and columns of the matrices in~\eqref{defgl} throughout this paper. So in~\eqref{defgl}, the elements of $A$ have negative indices only, those of $D$ have positive indices only, etc.
Furthermore, it will sometimes be convenient to write the minus sign of an index as an overlined number.
So with this convention the indices $\ldots, \bar{3},\bar{2},\bar{1};1,2,3,\ldots$ stand for $\ldots,-3,-2,-1;1,2,3\ldots$.

A basis for ${\mathfrak g}=\mathfrak{gl}(k|n)$ consists of matrices $E_{ij}$ ($i,j=-k,\ldots,-2,-1;1,2,\ldots,n$), 
with entry $1$ at position $(i,j)$ and $0$ elsewhere. This is usually called the Weyl basis.
The bracket for these basis elements is given by
\begin{equation}
\llbracket E_{ab}, E_{cd} \rrbracket = \delta_{bc} E_{ad} - (-1)^{\deg(E_{ab})\deg(E_{cd})} \delta_{ad} E_{cb}.
\label{defrel}
\end{equation}
A Cartan subalgebra ${\mathfrak h}$ of ${\mathfrak g}$ is spanned by the elements $E_{jj}$ 
($j=-k, \ldots,-2,-1;1,2,\ldots, n$).
The space dual to ${\mathfrak h}$ is ${\mathfrak h}^*$ and is described by the forms
$\epsilon_i$ ($i=-k, \ldots,-2,-1;1,2,\ldots,n$) where
$\epsilon_i:x\rightarrow A_{ii}$ ($i=-k, -k+1, \ldots, -1$) 
and $\epsilon_{i}:x\rightarrow D_{ii}$ ($i=1,2,\ldots, n$), and where $x$ is given as in (\ref{defgl}). 
The components of an element $\Lambda\in {\mathfrak h}^*$ will be written as 
\begin{align}
[m]^r& =[m_{-k,r},\ldots,m_{-2,r},m_{-1,r};m_{1r},m_{2r}, \ldots,m_{nr}]\nonumber\\
& = [m_{\bar k,r},\ldots,m_{\bar 2,r},m_{\bar 1,r};m_{1r},m_{2r}, \ldots,m_{nr}],
\label{components}
\end{align}
where 
\begin{equation}
\Lambda=\displaystyle \sum_{\myatop{\scriptstyle i=- k}{\scriptstyle i\ne 0}}^n m_{ir}\epsilon_i,
\end{equation} 
and $m_{ir}$ are complex numbers.
The elements of ${\mathfrak h}^*$ are called the weights. 
The roots of $\mathfrak{gl}(k|n)$ take the
form $\epsilon_i-\epsilon_j$ ($i\ne j$) and
the positive roots are $\epsilon_i-\epsilon_j$ ($i<j$).
By the index convention, $\epsilon_i-\epsilon_j$ is an even root when $i\cdot j>0$ and an odd root when $i\cdot j<0$.
So the positive odd roots are of the form $\epsilon_i-\epsilon_j$ with $i<0$ and $j>0$.
The distinguished set of simple roots is usually taken to be
\begin{equation}
\epsilon_{-k}-\epsilon_{-k+1},\ \epsilon_{-k+1}-\epsilon_{-k+2},\ldots, \epsilon_{- 1}-\epsilon_1,\  
 \epsilon_1-\epsilon_2,\ldots, \epsilon_{n-1}-\epsilon_n,
\label{simple}
\end{equation}
and this set contains only one odd simple root.
An element $\Lambda\in {\mathfrak h}^*$ with components $[m]^r$ will be called an integral dominant
weight if $m_{ir}-m_{i+1,r}\in{\mathbb Z}_+=\{0,1,2,\ldots\}$ for all $i\ne -1,0$
($-k\leq i\leq n-1$). 

For every integral dominant weight $\Lambda\equiv[m]^r$ there exists a unique finite-dimensional irreducible $\mathfrak{gl}(k|n)$ module
$V([\Lambda])$ with highest weight $\Lambda$.
The class of representations considered here is more restrictive.
We will concentrate on the covariant tensor representations (simply referred to as covariant representations), 
which are known to be finite-dimensional, irreducible and unitary.
Berele and Regev~\cite{BR} showed that the tensor product $V([1,0,\ldots, 0])^{\otimes N}$ of $N$ copies of the natural $(k+n)$-dimensional representation $V([1,0,\ldots, 0])$ of $\mathfrak{gl}(k|n)$ is completely reducible, and the irreducible components, $V_\lambda$,
can be labeled by a partition $\lambda$ of $N$ of length $l(\lambda)$ and weight $|\lambda|$, where 
$\lambda =(\lambda_1, \lambda_2, \ldots, \lambda_\ell)$, with $l(\lambda)=\ell$, $|\lambda|=\lambda_1+\lambda_2+\ldots+
\lambda_\ell=N$, and $\lambda_i\geq \lambda_{i+1}>0$ for $i=1,2,\ldots,\ell-1$, satisfying the condition $\lambda_{k+1}\leq n$.
For definitions regarding partitions, see~\cite{Macdonald}. 
The condition $\lambda_{k+1}\leq n$ is known as the {\em hook condition}: in terms of Young diagrams,
it means that the diagram of $\lambda$ should be inside the $(k,n)$-hook~\cite{BR}.

Covariant representations are also highest weight representations, so to every such partition $\lambda$ there should correspond an integral dominant weight $\Lambda^\lambda$ such that $V_{\lambda}$ is isomorphic to the highest weight module $V([\Lambda^\lambda])$. 
The relation between $\Lambda^\lambda$ $\equiv [m ]^r$ (see~\eqref{components})
and $\lambda =(\lambda_1, \lambda_2,\ldots)$ is such that~\cite{JHKR}: 
\begin{align}
& m_{ir}=\lambda_{k+i+1}, \quad -k\leq i\leq -1, \label{hwpart1}\\
& m_{ir}=\max\{0, \lambda'_i-k\}, \quad 1\leq i\leq n, \label{hwpart2}
\end{align}
where 
$\lambda'$ is the partition conjugate~\cite{Macdonald} to $\lambda$. 
Conversely if  $\Lambda$ $\equiv [m ]^r$ is integral dominant 
with all $m_{ir}\in{\mathbb Z}_+$ and 
\begin{equation}
m_{-1,r}\geq \# \{i:m_{ir}>0,\; 1\leq i \leq n\},
\label{geq}
\end{equation} 
then there exists a partition $\lambda$ such that $V([\Lambda])$ is isomorphic to the irreducible covariant tensor module
$V_\lambda$, and the components of $\lambda$ are given explicitly by
\begin{align}
& \lambda_{i}=m_{i-k-1,r}, \quad 1\leq i\leq k, \label{hwpart3}\\
& \lambda_{k+i}=\#\{j: m_{jr}\leq i, \;\; 1\leq j \leq n\}, \quad 1\leq i\leq n.\label{hwpart4}
\end{align}
Note that~\eqref{geq} can also be written in an alternative form:
\begin{equation}
\hbox{if }m_{-1,r}=p \hbox{ then }m_{ir}=0 \hbox{ for all } i>p.
\label{geq2}
\end{equation}

The main property of irreducible covariant tensor modules of $\mathfrak{gl}(k|n)$ is that their characters are known explicitly~\cite{BR, Sergeev}:
they are given by supersymmetric Schur functions. 
Explicitly, with the common notation
\[
x_{i}=e^{\epsilon_i} \quad(-k\leq i \leq -1),\qquad
y_i=e^{\epsilon_{i}} \quad(1\leq i \leq n),
\] 
one has
\begin{equation}
\mathop{\rm char}\nolimits V([\Lambda^\lambda]) = s_\lambda(x_{\bar k},\ldots,x_{\bar 2}, x_{\bar 1}|y_1, y_2,\ldots, y_n),
\end{equation} 
with $s_\lambda$ a supersymmetric Schur function~\cite{BR, Sergeev, King, Macdonald}.

In this paper, the decomposition of covariant representations of $\mathfrak{gl}(k|n)$ according to 
$\mathfrak{gl}(k|n) \supset \mathfrak{gl}(k-1|n)$ or to $\mathfrak{gl}(k|n) \supset \mathfrak{gl}(k|n-1)$ will be crucial for 
the construction of a proper Gel'fand-Zetlin basis.
Such decompositions, which are multiplicity-free, follow from properties of supersymmetric Schur functions~\cite{King, SV2010}.
In particular,
\begin{equation}
s_\lambda(x_{\bar k},x_{\overbar{k-1}},\ldots,x_{\bar 2}, x_{\bar 1}|y_1, y_2,\ldots, y_n) = 
\sum_{\sigma} x_{\bar k}^{|\lambda|-|\sigma|} s_\sigma(x_{\overbar{k-1}},\ldots,x_{\bar 2}, x_{\bar 1}|y_1, y_2,\ldots, y_n ),
\label{s-reduction}
\end{equation} 
where the sum is over all partitions $\sigma$ in the $(k-1,n)$-hook such that
\begin{equation}
\lambda_1 \geq \sigma_1 \geq \lambda_2 \geq \sigma_2 \geq \cdots
\geq \sigma_{l-1} \geq \lambda_l ,
\label{ineq-s}
\end{equation}
where $l$ stands for the length of $\lambda$.
In terms of the notions introduced in~\cite{Macdonald}, \eqref{ineq-s} means that $\lambda-\sigma$ is
a {\em horizontal strip}.
Similarly,
\begin{equation}
s_\lambda(x_{\bar k},\ldots,x_{\bar 2}, x_{\bar 1}|y_1,\ldots,y_{n-1},y_n) = 
\sum_{\sigma} s_\sigma(x_{\bar k},\ldots,x_{\bar 2}, x_{\bar 1}|y_1,\ldots,y_{n-1}) \cdot y_n^{|\lambda|-|\sigma|},
\label{ss-reduction}
\end{equation} 
where the sum is now over all partitions $\sigma$ in the $(k,n-1)$-hook such that
\begin{equation}
\lambda_1' \geq \sigma_1' \geq \lambda_2' \geq \sigma_2' \geq \cdots
\geq \sigma_{\ell-1}' \geq \lambda_\ell'\;,
\label{ineq-ss}
\end{equation}
where $\ell=\lambda_1$ stands for the length of $\lambda'$.
Following the terminology of~\cite{Macdonald}, $\lambda-\sigma$ is a {\em vertical strip}.

\setcounter{equation}{0}
\section{The odd GZ-basis for covariant representations of $\mathfrak{gl}(n|n)$} 
\label{sec:representations} 

The purpose of this section is to construct a new GZ-basis for covariant representations of $\mathfrak{gl}(n|n)$. 
In this construction, the labelling arising from the decomposition of the representation according to the subalgebra chain
\[
\mathfrak{gl}(n|n) \supset \mathfrak{gl}(n|n-1) \supset \mathfrak{gl}(n-1|n-1) 
\]
will be used. 
For the decomposition itself, it is clear that the Schur function identities of the previous section give all necessary information.
The second part of this section is devoted to computing the action of a set of $\mathfrak{gl}(n|n)$ generators on the GZ-basis vectors.

Let $V([m]^{r})$, $r=2n$ be an irreducible covariant tensor module of $\mathfrak{gl}(n|n)$ 
determined by its highest weight, given by the nonnegative integer $r$-tuple  
\begin{equation}
[m]^{r}=[m_{-n,r}, \ldots,m_{-2,r} ,m_{-1,r};m_{1r},m_{2r}, \ldots, m_{nr}],  \label{mr}
\end{equation}
such that
\begin{equation}
m_{ir}-m_{i+1,r}\in {\mathbb Z}_+, \; 
 \; i =-n,\ldots,-2;1,\ldots, n-1 \label{cond1}
\end{equation}
and 
\begin{equation}
m_{-1,r}\geq \# \{i:m_{ir}>0,\; 1\leq i \leq n\}. \label{cond2}
\end{equation}
Within a given $\mathfrak{gl}(n|n)$ module $V([m]^{r})$ the numbers~(\ref{mr}) are fixed.

\begin{prop}
\label{prop1}
Consider the $\mathfrak{gl}(n|n)$ module $V([m]^{r})$ as a $\mathfrak{gl}(n|n-1)$ module.
Then $V([m]^{r})$ can be represented as a direct sum of covariant simple $\mathfrak{gl}(n|n-1)$ modules,
\begin{equation}
V([m]^{r})=\sum_i \oplus V_i([m]^{r-1}), \label{gl(mn-1)}
\end{equation}
where 
\begin{itemize}
\item[I.] All $V_i([m]^{r-1})$ carry inequivalent representations of $\mathfrak{gl}(n|n-1)$
\begin{align*}
& [m]^{r-1}=[m_{-n,r-1}, \ldots, m_{-1,r-1};m_{1,r-1},\ldots , m_{n-1,r-1}], \nonumber\\
& m_{i,r-1}-m_{i+1,r-1}\in{\mathbb Z}_+,\;
i =-n,\ldots,-2,1,\ldots, n-2,\\
&  m_{-1,r-1}\geq \# \{i:m_{i,r-1}>0,\; 1\leq i \leq n-1\} .
\end{align*}
\item[II.] 
\begin{equation}
 \begin{array}{rl}
1.& m_{ir}-m_{i,r-1}=\theta_{i,r-1}\in\{0,1\},\quad -n\leq i\leq -1,\\
2.& m_{ir}-m_{i,r-1}\in{\mathbb Z}_+\hbox{ and }\; m_{i,r-1}-m_{i+1,r}\in{\mathbb Z}_+,\quad
    1\leq i\leq n-1.
 \end{array}
\label{cond0}
\end{equation}
\end{itemize}
\end{prop}

\noindent {\bf Proof.} The above conditions follow from~\eqref{ss-reduction}, with $\lambda$ the partition corresponding to the $\mathfrak{gl}(n|n)$ highest weight $[m]^r$. 
The conditions~I come from the fact that the partition $\sigma$ in~\eqref{ss-reduction} determines a covariant representation of $\mathfrak{gl}(n|n-1)$, with highest weight denoted by $[m]^{r-1}$. 
The conditions~II follow from~\eqref{ineq-ss}, or from the fact that $\lambda-\sigma$ is a vertical strip. When these ``interlacing'' conditions for $\lambda$ and $\sigma$ are rewritten in terms of the corresponding highest weights $[m]^r$ and $[m]^{r-1}$, \eqref{cond0} arises.
$\phantom{some text}$ \mybox

\begin{prop}
\label{prop2}
Consider a covariant $\mathfrak{gl}(n|n-1)$ module $V([m]^{r-1})$ as a $\mathfrak{gl}(n-1|n-1)$ module.
Then $V([m]^{r-1})$ can be represented as a direct sum of simple $\mathfrak{gl}(n-1|n-1)$ modules $V([m]^{r-2})$,
\begin{equation}
V([m]^{r-1})=\sum_i \oplus V_i([m]^{r-2}), \label{gl(m)}
\end{equation}
where 
\begin{itemize}
\item[I.] All $V_i([m]^{r-2})$ carry inequivalent representations of $\mathfrak{gl}(n-1|n-1)$
\begin{align*}
& [m]^{r-2}=[m_{-n+1,r-2}, \ldots, m_{-1,r-2};m_{1,r-2},\ldots , m_{n-1,r-2}], \nonumber\\
& m_{i,r-2}-m_{i+1,r-2}\in{\mathbb Z}_+,\;
 i =-n+1,\ldots,-2,1,\ldots, n-2,\\
&  m_{-1,r-2}\geq \# \{i:m_{i,r-2}>0,\; 1\leq i \leq n-1\} .
\end{align*}
\item[II.] 
\begin{equation}
 \begin{array}{rl}
1.& m_{i,r-2}-m_{i,r-1}=\theta_{i,r-2}\in\{0,1\},\quad 1\leq i\leq n-1,\\
2.& m_{i,r-1}-m_{i+1,r-2}\in{\mathbb Z}_+\hbox{ and }\; m_{i+1,r-2}-m_{i+1,r-1}\in{\mathbb Z}_+,\quad
    -n\leq i\leq -2.
 \end{array}
\label{cond}
\end{equation}
\end{itemize}
\end{prop}

\noindent {\bf Proof.} 
The proof is the same as for the previous proposition, but now we use~\eqref{s-reduction}.
The ``interlacing'' conditions~\eqref{ineq-s} for $\lambda$ and $\sigma$, rewritten in terms of the corresponding highest weights $[m]^{r-1}$ and $[m]^{r-2}$, give rise to~\eqref{cond}. \mybox

Since in each step of the reduction
\begin{equation}
\mathfrak{gl}(n|n) \supset \mathfrak{gl}(n|n-1) \supset \mathfrak{gl}(n-1|n-1) \supset \mathfrak{gl}(n-1|n-2) \supset \mathfrak{gl}(n-2|n-2) \supset 
\cdots \supset \mathfrak{gl}(1|1) \supset \mathfrak{gl}(1) 
\label{chain}
\end{equation}
the decomposition of $V([m]^r)$  is multiplicity-free and follows simple rules, this gives rise to an interesting GZ-basis for the basis vectors
of $V([m]^r)$.
Hence Propositions~\ref{prop1} and~\ref{prop2}, and the fact that representations of $\mathfrak{gl}(1)$ are one-dimensional, imply:
 
\begin{prop}
\label{prop3}
The set of vectors \\
\noindent
 $|m)^{r} =$ 
\begin{equation}
 \left|
\begin{array}{llcll:llclll}
m_{\bar n r} & m_{\overbar{n-1},r} & \cdots & m_{\bar 2r} & m_{\bar 1r} & m_{1r} & m_{2r} &\cdots & m_{n-2,r} &m_{n-1,r} &m_{nr}\\
 \uparrow & \uparrow & \cdots & \uparrow &\uparrow & &&&&&\\
m_{\bar n, r-1} & m_{\overbar{n-1}, r-1} & \cdots & m_{\bar 2, r-1} & m_{\bar 1, r-1} & m_{1,r-1} & m_{2,r-1} &\cdots &m_{n-2,r-1} & m_{n-1,r-1} &\\
&&&&&\downarrow & \downarrow & \cdots & \downarrow &\downarrow  \\
 & m_{\overbar{n-1},r-2} & \cdots & m_{\bar 2,r-2} & m_{\bar 1,r-2} & m_{1,r-2} & m_{2,r-2} &\cdots & m_{n-2,r-2} & m_{n-1,r-2} &\\
 &\uparrow & \cdots & \uparrow &\uparrow &&&&\\
 & m_{\overbar{n-1},r-3} & \cdots & m_{\bar 2,r-3} & m_{\bar 1,r-3} & m_{1,r-3} & m_{2,r-3} &\cdots & m_{n-2,r-3}  &  &\\
 &  &\ddots &\vdots & \vdots & \vdots &\vdots &\iddots & & \\
&&& m_{\bar 2 4} &  m_{\bar 1 4} & m_{14} & m_{24}& & & & \\
&&&\uparrow & \uparrow \\
&&& m_{\bar 2 3} &  m_{\bar 1 3} & m_{13} && & & & \\
&&&&&\downarrow \\
&&&& m_{\bar 1 2} & m_{12} & & & & & \\
&&&& \uparrow\\
&&&& m_{\bar 1 1}  & & & & & &
\end{array}
\right)
\label{mn}
\end{equation}
satisfying the conditions
\begin{equation}
 \begin{array}{rl}
1. & m_{ir}\in{\mathbb Z}_+ \; \hbox{are fixed and }  m_{jr}-m_{j+1,r}\in{\mathbb Z}_+ , \;j\neq -1,0,\;
    -n\leq j\leq n-1,\\
    & m_{-1,r}\geq \# \{i:m_{ir}>0,\; 1\leq i \leq n\};\\
2.& m_{i,2p}-m_{i,2p-1}\equiv\theta_{i,2p-1}\in\{0,1\},\quad 1\leq p\leq n;\;
    -p\leq i\leq -1;\\
3.& m_{i,2p}-m_{i,2p+1}\equiv\theta_{i,2p}\in\{0,1\},\quad 1\leq p\leq n-1;\;
    1\leq i\leq p;\\    
4.  &   m_{-1,2p}\geq \# \{i:m_{i,2p}>0,\; 1\leq p \leq n, \;  1\leq i\leq p\} ;\\  
5.  &   m_{-1,2p-1}\geq \# \{i:m_{i,2p-1}>0,\; 2\leq p \leq n, \;  1\leq i\leq p-1\} ;\\ 
 
6.& m_{i,2p}-m_{i,2p-1}\in{\mathbb Z}_+\hbox{ and } m_{i,2p-1}-m_{i+1,2p}\in{\mathbb Z}_+,\\
  &  2\leq p\leq n, \quad   1\leq i\leq  p-1;\\
7.& m_{i,2p+1}-m_{i+1,2p}\in{\mathbb Z}_+\hbox{ and } m_{i+1,2p}-m_{i+1,2p+1}\in{\mathbb Z}_+,\\
  &  1\leq p\leq n-1,  \quad -p-1\leq i\leq  -2. 
 \end{array}
\label{cond3}
\end{equation}
constitute a basis in $V([m]^{r})$.
\end{prop}

Conditions~2 and~3 are referred to as ``$\theta$-conditions''.
Conditions~6 and~7 are often referred to as ``in-betweenness conditions.''
Note that the arrows in this pattern have no real function, and can be omitted. 
We find it useful to include them, just in order to visualize the conditions.
When there is an arrow $a\rightarrow b$ between labels $a$ and $b$, it means that either $b=a$ or else $b=a+1$ (a $\theta$-condition).
We will also refer to ``rows'' and ``columns'' of the GZ-pattern.
Rows are counted from the bottom: row~1 is the bottom row in~\eqref{mn}, and row~$r$ is the top row in~\eqref{mn}.
In an obvious way, columns~$1$, $2$, $3\cdots$ refer to the columns to the right of the dashed line in~\eqref{mn},
and columns~$-1$, $-2$, $-3,\cdots$ (or $\bar 1$, $\bar 2$, $\bar 3,\cdots$) to the columns to the left of this dashed line.
For two consecutive rows in the GZ-pattern~\eqref{mn}, about half of the labels involve $\theta$-conditions, and the other half involves in-betweenness conditions.
Note that a row in the GZ-pattern always corresponds to the highest weight labels of some subalgebra $\mathfrak{gl}(k|l)$ in the chain~\eqref{chain}.
The dashed line in~\eqref{mn} also has no function, but is there just to distinguish between the labels on the left (corresponding to $\mathfrak{gl}(k)$) and those to the right (corresponding to $\mathfrak{gl}(l)$).

We shall refer to the basis~(\ref{mn}) as the odd GZ-basis for the covariant $\mathfrak{gl}(n|n)$ representations.
The task is now to give the explicit action of a set of $\mathfrak{gl}(n|n)$  generators 
on the basis vectors~(\ref{mn}). 

When describing actions of a Lie algebra or Lie superalgebra on some GZ-basis vector, this is usually not done for all algebra basis elements but only for a set of generators. The actions of the remaining algebra elements then follows by applying the Lie (super)algebra brackets.
For the case of $\mathfrak{gl}(n|n)$, if one works in the traditional GZ-basis~\cite{SV2010}, one is describing the action of the following generators:
\begin{itemize}
\item the basis elements $E_{ii}$ ($i=-n,\ldots,-1;1,\ldots, n$) of the Cartan subalgebra,
\item the Chevalley generators $E_{-1,1},\;E_{i,i+1}$ ($i=-n,\ldots,-2; 1, \ldots, n-1$), corresponding to the set of simple roots~\eqref{simple},
\item the Chevalley generators $E_{1,-1}, \;E_{i+1,i}$ ($i=-n,\ldots,-2; 1, \ldots, n-1$), corresponding to the negatives of the simple roots.
\end{itemize}
These form indeed a set of generating elements.
The reason for this choice of generators is the following: amongst all basis elements $E_{ij}$ with $i\ne j$, the above Chevalley generators have the ``simplest'' action on a traditional GZ-basis element~\cite{SV2010}. This simplicity is evident from the fact that by the action of the generators the GZ-labels change only in one row of the GZ-pattern.

Currently, we are dealing with a different GZ-basis, the odd GZ-basis for $\mathfrak{gl}(n|n)$.
So it is not surprising that the choice of generators, in order to describe the action on the GZ-basis elements, is different.
Of course, the ``diagonal action'' by means of the Cartan subalgebra basis elements will still remain.
For the other generators, it is natural to make the following choice:
\begin{itemize}
\item the positive root vectors
\begin{equation}
E_{-1,1}, E_{-2,1}, E_{-2,2}, E_{-3,2}, E_{-3,3}, \ldots, E_{-n,n-1}, E_{-n,n},
\label{posE}
\end{equation}
\item the negative root vectors
\begin{equation}
E_{1,-1}, E_{1,-2}, E_{2,-2}, E_{2,-3}, E_{3,-3}, \ldots, E_{n-1,-n}, E_{n,-n}.
\label{negE}
\end{equation}
\end{itemize}
Note that these would correspond to the root vectors $E_{\pm\alpha}$ for the following choice of simple roots:
\begin{equation}
\epsilon_{-1}-\epsilon_1,\ \epsilon_1-\epsilon_{-2},\ \epsilon_{-2}-\epsilon_2,\ \epsilon_2-\epsilon_{-3},\ldots,
\epsilon_{-n+1}-\epsilon_{n-1},\ \epsilon_{n-1}-\epsilon_{-n},\ \epsilon_{-n}-\epsilon_n.
\label{oddsimple}
\end{equation}
This set of simple roots contains odd roots only, which is the reason to refer to our new GZ-basis as ``the odd GZ-basis''.
From the relation with this set of simple roots~\eqref{oddsimple}, it also follows that the elements~\eqref{posE}-\eqref{negE} (together with the $E_{ii}$) form indeed a set of generators.

The following theorem is one of the two main results of this paper.
It describes the action of the new set of generators on the odd GZ-basis vectors.
In this theorem we use some further notation ($\varphi =0,1; \; k=1,\ldots,n$):
\begin{equation}
l_{i,2k-\varphi}=m_{i,2k-\varphi}-i \quad(-k\leq i \leq -1);\qquad
l_{p,2k-\varphi}=-m_{p,2k-\varphi}+p \quad(1\leq p\leq k-\varphi).
\label{lir}
\end{equation}
Furthermore, $|m)^r_{\pm (ij)}$ is the pattern obtained
from $|m)^r$ by replacing the entry $m_{ij}$ by $m_{ij}\pm 1$.

\begin{theo} 
\label{theo4}
The transformation of the irreducible covariant tensor
module $V([m]^{r})$ under the action of the $\mathfrak{gl}(n|n)$ generators is given by:

\begin{align}
& E_{-i,-i}|m)^r=\left(\sum_{j\neq 0=-i}^{i-1} m_{j,2i-1}-\sum_{j\neq 0=-i+1}^{i-1} m_{j,2i-2}\right)|m)^r, 
\quad 1\leq i\leq n; \label{Ekk}\\
& E_{ii}|m)^r=\left(\sum_{j\neq 0 =-i}^{i} m_{j,2i}-\sum_{j\neq 0=-i}^{i-1} m_{j,2i-1}\right)|m)^r, 
\quad 1\leq i\leq n; \label{E_nk}
\end{align}
\begin{align}
& E_{-i,i}|m)^r=\sum_{k=-i}^{-1} \theta_{k,2i-1}
S\Bigl({\sum_{s=1}^{i-1}\sum_{j=-s}^{-1}\theta_{j,2s-1}+\sum_{s=1}^{i-1}\sum_{j=1}^{s}\theta_{j,2s}+\sum_{j=-i}^{k-1}\theta_{j,2i-1}}\Bigr) \nonumber\\
&\times \left(\frac{
{\prod_{j=-i+1}^{-1} (l_{k,2i}-l_{j,2i-2}-1)
\prod_{j=1}^{i} (l_{k,2i}-l_{j,2i})}}
{\prod_{j\neq k=-i}^{-1} (l_{k,2i}-l_{j,2i}) \prod_{j=1}^{i-1}
(l_{k,2i}-l_{j,2i-2}-1) } \right)^{1/2}|m)^r_{+(k,2i-1)}\; \nonumber\\
& +\sum_{k=1}^{i-1} \theta_{k,2i-2}
S\Bigl({\sum_{s=1}^{i-1}\sum_{j=-s}^{-1}\theta_{j,2s-1}+\sum_{s=1}^{i-2}\sum_{j=1}^{s}\theta_{j,2s}+\sum_{j=1}^{k-1}\theta_{j,2i-2}}\Bigr) \nonumber\\
&\times \left(-\frac{
{\prod_{j=-i+1}^{-1} (l_{j,2i-2}-l_{k,2i-1}+1)
\prod_{j=1}^{i} (l_{j,2i}-l_{k,2i-1})}}
{\prod_{j=-i}^{-1} (l_{j,2i}-l_{k,2i-1})\prod_{j\neq k=1}^{i-1}
(l_{j,2i-2}-l_{k,2i-1}+1) } \right)^{1/2}|m)^r_{+(k,2i-1)},\; \nonumber\\
& \hskip 6cm 1\leq i\leq n;\label{Eni}
\end{align}
\begin{align}
& E_{-i-1,i}|m)^r=\sum_{k=-i}^{-1} \theta_{k,2i-1}
S\Bigl({\sum_{s=1}^{i-1}\sum_{j=-s}^{-1}\theta_{j,2s-1}+\sum_{s=1}^{i-1}\sum_{j=1}^{s}\theta_{j,2s}+\sum_{j=-i}^{k-1}\theta_{j,2i-1}}\Bigr) \nonumber\\
&\times \left(-\frac{
{\prod_{j=-i-1}^{-1} (l_{k,2i}-l_{j,2i+1})
\prod_{j=1}^{i-1} (l_{k,2i-1}-l_{j,2i-1})}}
{\prod_{j\neq k=-i}^{-1} (l_{k,2i-1}-l_{j,2i-1})\prod_{j=1}^{i}
(l_{k,2i-1}-l_{j,2i+1}+1) } \right)^{1/2}|m)^r_{-(k,2i)}\; \nonumber\\
& +\sum_{k=1}^{i} \theta_{k,2i}
S\Bigl({\sum_{s=1}^{i}\sum_{j=-s}^{-1}\theta_{j,2s-1}+\sum_{s=1}^{i-1}\sum_{j=1}^{s}\theta_{j,2s}+\sum_{j=1}^{k-1}\theta_{j,2i}}\Bigr) \nonumber\\
&\times \left(\frac{
{\prod_{j=-i-1}^{-1} (l_{j,2i+1}-l_{k,2i+1})
\prod_{j=1}^{i-1} (l_{j,2i-1}-l_{k,2i})}}
{\prod_{j=-i}^{-1} (l_{j,2i-1}-l_{k,2i+1}+1)\prod_{j\neq k=1}^{i}
(l_{j,2i+1}-l_{k,2i+1}) } \right)^{1/2}|m)^r_{-(k,2i)},\; \nonumber\\
& \hskip 6cm 1\leq i\leq n-1;\label{Eni1}
\end{align}
\begin{align}
& E_{i,-i}|m)^r=\sum_{k=-i}^{-1} (1-\theta_{k,2i-1})
S\Bigl({\sum_{s=1}^{i-1}\sum_{j=-s}^{-1}\theta_{j,2s-1}+\sum_{s=1}^{i-1}\sum_{j=1}^{s}\theta_{j,2s}+\sum_{j=-i}^{k-1}\theta_{j,2i-1}}\Bigr) \nonumber\\
&\times \left(\frac{
{\prod_{j=-i+1}^{-1} (l_{k,2i}-l_{j,2i-2}-1)
\prod_{j=1}^{i} (l_{k,2i}-l_{j,2i})}}
{\prod_{j\neq k=-i}^{-1} (l_{k,2i}-l_{j,2i}) \prod_{j=1}^{i-1}
(l_{k,2i}-l_{j,2i-2}-1) } \right)^{1/2}|m)^r_{-(k,2i-1)}\; \nonumber\\
& +\sum_{k=1}^{i-1} (1-\theta_{k,2i-2})
S\Bigl({\sum_{s=1}^{i-1}\sum_{j=-s}^{-1}\theta_{j,2s-1}+\sum_{s=1}^{i-2}\sum_{j=1}^{s}\theta_{j,2s}+\sum_{j=1}^{k-1}\theta_{j,2i-2}}\Bigr) \nonumber\\
&\times \left(-\frac{
{\prod_{j=-i+1}^{-1} (l_{j,2i-2}-l_{k,2i-1})
\prod_{j=1}^{i} (l_{j,2i}-l_{k,2i-1}-1)}}
{\prod_{j=-i}^{-1} (l_{j,2i}-l_{k,2i-1}-1)\prod_{j\neq k=1}^{i-1}
(l_{j,2i-2}-l_{k,2i-1}) } \right)^{1/2}|m)^r_{-(k,2i-1)},\; \nonumber\\
& \hskip 6cm 1\leq i\leq n;\label{Fni}
\end{align}
\begin{align}
& E_{i,-i-1}|m)^r=\sum_{k=-i}^{-1} (1-\theta_{k,2i-1})
S\Bigl({\sum_{s=1}^{i-1}\sum_{j=-s}^{-1}\theta_{j,2s-1}+\sum_{s=1}^{i-1}\sum_{j=1}^{s}\theta_{j,2s}+\sum_{j=-i}^{k-1}\theta_{j,2i-1}}\Bigr) \nonumber\\
&\times \left(-\frac{
{\prod_{j=-i-1}^{-1} (l_{k,2i}-l_{j,2i+1}+1)
\prod_{j=1}^{i-1} (l_{k,2i-1}-l_{j,2i-1})}}
{\prod_{j\neq k=-i}^{-1} (l_{k,2i-1}-l_{j,2i-1})\prod_{j=1}^{i}
(l_{k,2i-1}-l_{j,2i+1}+1) } \right)^{1/2}|m)^r_{+(k,2i)}\; \nonumber\\
& +\sum_{k=1}^{i} (1-\theta_{k,2i})
S\Bigl({\sum_{s=1}^{i}\sum_{j=-s}^{-1}\theta_{j,2s-1}+\sum_{s=1}^{i-1}\sum_{j=1}^{s}\theta_{j,2s}+\sum_{j=1}^{k-1}\theta_{j,2i}}\Bigr) \nonumber\\
&\times \left(\frac{
{\prod_{j=-i-1}^{-1} (l_{j,2i+1}-l_{k,2i+1})
\prod_{j=1}^{i-1} (l_{j,2i-1}-l_{k,2i}+1)}}
{\prod_{j=-i}^{-1} (l_{j,2i-1}-l_{k,2i+1}+1)\prod_{j\neq k=1}^{i}
(l_{j,2i+1}-l_{k,2i+1}) } \right)^{1/2}|m)^r_{+(k,2i)},\; \nonumber\\
& \hskip 6cm 1\leq i\leq n-1.\label{Fni1}
\end{align}
\end{theo}

In the above expressions, $\sum_{j\ne l=m}^p$ means that
$j$ takes all values from $m$ to $p$ with $j\ne l$. 
We have also used the abbreviation $S(a)=(-1)^a$ when the expression $a$ is lengthy.

All vectors $|m)^r$ of the module $V([m]^r)$ are weight vectors, so the action of the elements $E_{ii}$ on $|m)^r$ is diagonal.
In fact, \eqref{Ekk} and~\eqref{E_nk} just yield the weight of the corresponding vector.
The action of the other generators on $|m)^r$ is ``simple'', in the sense that the action changes only one label of the GZ-pattern by $+1$ or $-1$, and all changes appear in one row of the pattern only.

We shall only give an outline of the proof of Theorem~\ref{theo4}, since it is very technical.
The main part of the proof consist of showing that the composition of the actions~\eqref{Eni}-\eqref{Fni1} by means of (anti)commutators yield the correct action. 
So one has to show that
\begin{align}
& (E_{-i,i} E_{i,-i} +  E_{i,-i} E_{-i,i}) |m)^r = (E_{-i,-i} + E_{ii}) |m)^r, \; 1\leq i\leq n,\label{EE1}\\
& (E_{-i-1,i} E_{i,-i-1} +  E_{i,-i-1} E_{-i-1,i}) |m)^r = (E_{-i-1,-i-1} + E_{ii}) |m)^r, \;1\leq i\leq n-1. \label{EE2}
\end{align}
In such computations, the left hand side becomes extremely complicated, whereas the right hand side is simple.
The simplification of the left hand side involves several steps, and it is instructive to explain these steps by means of a generic example.
It should be clear that the arguments used in the example hold in general.

So let us take for a moment $n=4$, $i=2$, and consider the action of the left hand side of~\eqref{EE2} on a general vector with GZ-labels ($r=8$)
\begin{equation}
|m)^8 =  \left|
\begin{array}{llll:llll}
m_{\bar 4 8} & m_{\bar 3 8} & m_{\bar 2 8} & m_{\bar 1 8} & m_{18} & m_{28} & m_{38} & m_{48} \\
m_{\bar 4 7} & m_{\bar 3 7} & m_{\bar 2 7} & m_{\bar 1 7} & m_{17} & m_{27} & m_{37} &  \\
 & m_{\bar 3 6} & m_{\bar 2 6} & m_{\bar 1 6} & m_{16} & m_{26} & m_{36} &  \\
 & m_{\bar 3 5} & m_{\bar 2 5} & m_{\bar 1 5} & m_{15} & m_{25} &  &  \\
 &  & m_{\bar 2 4} & m_{\bar 1 4} & m_{14} & m_{24} &  &  \\
 &  & m_{\bar 2 3} & m_{\bar 1 3} & m_{13} &  &  &  \\
 &  &  & m_{\bar 1 2} & m_{12} &  &  &  \\
 &  &  & m_{\bar 1 1} &  &  &  &  
\end{array}
\right).
\label{m8}
\end{equation}
In particular, because of the $\theta$-conditions, one can write this vector as
\begin{equation}
 \left|
\begin{array}{llll:llll}
m_{\bar 4 7}+\theta_{\bar 4 7} & m_{\bar 3 7}+\theta_{\bar 3 7} & m_{\bar 2 7}+\theta_{\bar 2 7} & m_{\bar 1 7}+\theta_{\bar 1 7} & m_{18} & m_{28} & m_{38} & m_{48} \\
m_{\bar 4 7} & m_{\bar 3 7} & m_{\bar 2 7} & m_{\bar 1 7} & m_{17} & m_{27} & m_{37} &  \\
 & m_{\bar 3 5}+\theta_{\bar 3 5} & m_{\bar 2 5}+\theta_{\bar 2 5} & m_{\bar 1 5}+\theta_{\bar 1 5} & m_{17}+\theta_{16} & m_{27}+\theta_{26} & m_{37}+\theta_{36} &  \\
 & m_{\bar 3 5} & m_{\bar 2 5} & m_{\bar 1 5} & m_{15} & m_{25} &  &  \\
 &  & m_{\bar 2 3}+\theta_{\bar 2 3} & m_{\bar 1 3}+\theta_{\bar 1 3} & m_{15}+\theta_{14} & m_{25}+\theta_{24} &  &  \\
 &  & m_{\bar 2 3} & m_{\bar 1 3} & m_{13} &  &  &  \\
 &  &  & m_{\bar 1 1}+\theta_{\bar 1 1} & m_{13}+\theta_{12} &  &  &  \\
 &  &  & m_{\bar 1 1} &  &  &  &  
\end{array}
\right),
\label{m8t}
\end{equation}
where essentially each $\theta$ can be 0 or 1.
Compute now the action $(E_{\bar 32}E_{2\bar 3}+E_{2\bar 3}E_{\bar 32}) |m)^8$ on the vector~\eqref{m8t}, using~\eqref{Eni1} and~\eqref{Fni1}, and collect the terms with the same GZ-pattern. 
It is fairly easy (but technical) to verify that all non-diagonal terms (i.e.\ the terms with a GZ-pattern that is different from $|m)^8$) cancel automatically because of the sign-factors $S(\cdot)$.
The coefficient of the diagonal term (i.e.\ the term in $|m)^8$ itself) yields, using the expressions~\eqref{Eni1} and~\eqref{Fni1}:
\begin{align}
&\frac{(m_{\bar 3 5}-m_{\bar 1 3}+1)(m_{\bar 2 5}-m_{\bar 1 3})(m_{\bar 1 3}-m_{\bar 1 5}+1)(m_{\bar 1 3}+m_{13})}
{(m_{\bar 1 3}+m_{15}+1)(m_{\bar 1 3}+m_{25})(m_{\bar 2 3}-m_{\bar 1 3}+1)}\nonumber\\
&+\frac{(m_{\bar 3 5}+m_{15}+2)(m_{\bar 2 5}+m_{15}+1)(m_{\bar 1 5}+m_{15})(m_{15}-m_{13}+1)}
{(m_{\bar 2 3}+m_{15}+2)(m_{\bar 1 3}+m_{15}+1)(m_{15}-m_{25}+1)}\nonumber\\
&+\frac{(m_{\bar 3 5}+m_{25}+1)(m_{\bar 2 5}+m_{25})(m_{\bar 1 5}+m_{25}-1)(m_{13}-m_{25})}
{(m_{\bar 2 3}+m_{25}+1)(m_{\bar 1 3}+m_{25})(m_{15}-m_{25}+1)}\nonumber\\
&+\frac{(m_{\bar 3 5}-m_{\bar 2 3})(m_{\bar 2 3}-m_{\bar 2 5}+1)(m_{\bar 2 3}-m_{\bar 1 5}+2)(m_{\bar 2 3}+m_{13}+1)}
{(m_{\bar 2 3}+m_{15}+2)(m_{\bar 2 3}+m_{25}+1)(m_{\bar 2 3}-m_{\bar 1 3}+1)}.
\label{coeff1}
\end{align}
Observe that this coefficient is independent of the $\theta$-values, so that reduces the number of cases to be considered.
In order to understand the structure of such coefficients, let us make the following substitutions in this last expression: 
\begin{align*}
& m_{\bar 2 3}=-x_0,\ m_{\bar 1 3}=-x_1+1,\ m_{15}=x_2-2,\ m_{25}=x_3-1,\\  
& m_{\bar 3 5}=-a_0,\ m_{\bar 2 5}=-a_1+1,\ m_{\bar 1 5}=-a_2+2,\ m_{13}=a_3-1.
\end{align*}
This yields:
\begin{align}
& \frac{(x_1-a_0)(x_1-a_1)(x_1-a_2)(x_1-a_3)}{(x_1-x_0)(x_1-x_2)(x_1-x_3)}+
\frac{(x_2-a_0)(x_2-a_1)(x_2-a_2)(x_2-a_3)}{(x_2-x_0)(x_2-x_1)(x_2-x_3)} \nonumber\\
& \frac{(x_3-a_0)(x_3-a_1)(x_3-a_2)(x_3-a_3)}{(x_3-x_0)(x_3-x_1)(x_3-x_2)}+
\frac{(x_0-a_0)(x_0-a_1)(x_0-a_2)(x_0-a_3)}{(x_0-x_1)(x_0-x_2)(x_0-x_3)}.
\label{coeff2}
\end{align}
This expression simplifies (see soon), and yields
\[
x_0+x_1+x_2+x_3-a_0-a_1-a_2-a_3 = m_{\bar 3 5}+m_{\bar 2 5}+m_{\bar 1 5}+m_{15}+m_{25}-m_{\bar 2 3}-m_{\bar 1 3}-m_{13}.
\]
With \eqref{Ekk}-\eqref{E_nk}, one also finds
\[
(E_{\bar 3 \bar 3}+E_{22}) |m)^8 = (m_{\bar 3 5}+m_{\bar 2 5}+m_{\bar 1 5}+m_{15}+m_{25}-m_{\bar 2 3}-m_{\bar 1 3}-m_{13}) |m)^8.
\]
So we have shown, for this example, that the anticommutator relation $\{E_{\bar 3 2},E_{2\bar 3}\}=E_{\bar 3 \bar 3}+E_{22}$ holds when acting on an arbitrary vector of the form~\eqref{m8}.

For the general verification of relations~\eqref{EE1}-\eqref{EE2}, the arguments are the same.
The non-diagonal terms arising in the left hand side cancel, and the coefficient of the diagonal term is of the form~\eqref{coeff1}.
After making appropriate substitutions, the coefficient always reduces to an expression of the form~\eqref{coeff2}. 
For the simplification of such expressions, one can make use of the following rational function identity:
\begin{equation}
\sum_{i=0}^n \ \frac{\displaystyle \prod_{j=0}^n (x_i-a_j)}{\displaystyle \prod_{\myatop{\scriptstyle j=0}{\scriptstyle j\ne i}}^n (x_i-x_j)} = \sum_{i=0}^n (x_i-a_i).
\label{lagrange-id}
\end{equation}
The proof of such identities is standard~\cite{Molev}, and uses the Lagrange interpolation formula. 
For completeness, we present a short proof in the Appendix.

So far, we have described the proof of~\eqref{EE1}-\eqref{EE2}, and this does not yet prove Theorem~\ref{theo4} completely.
To continue, one should consider the remaining basis elements of $\mathfrak{gl}(n|n)$ and their action on $|m)^r$.
The action of other basis elements is determined by consecutive actions of the generators, e.g.\
\begin{align*}
& E_{-n, -n+1}|m)^r= (E_{-n,n-1}E_{n-1,-n+1}+E_{n-1,-n+1}E_{-n,n-1}) |m)^r,\\
& E_{-n+1,-n+2}|m)^r= (E_{-n+1,n-2}E_{n-2,-n+2}+E_{n-2,-n+2}E_{-n+1,n-2}) |m)^r, \hbox{ etc.}
\end{align*}
Once this has been achieved, it remains to verify that the defining relations~\eqref{defrel} hold when acting on an arbitrary vector $|m)^r$.
This, however, is easier than it sounds: 
the validity of the corresponding equation follows from consecutive use of the Jacobi-identity, from the definition in terms of the generators themselves, from the equations~\eqref{EE1}-\eqref{EE2}, or from simple cancellations due to the sign factors $S(\cdot)$ 
in~\eqref{Eni}-\eqref{Fni1}.
So this completes the outline of the proof of Theorem~\ref{theo4}. \mybox

We end this section by some further remarks.
First of all, note that the action of the original Chevalley generators corresponding to the positive even simple roots follows from
($i=1,\ldots,n-1$)
\begin{align}
& E_{i,i+1}|m)^r= (E_{i,-i-1}E_{-i-1,i+1}+E_{-i-1,i+1}E_{i,-i-1}) |m)^r,\nonumber\\
& E_{-i-1,-i}|m)^r= (E_{-i-1,i}E_{i,-i}+E_{i,-i}E_{-i-1,i}) |m)^r. \label{Chev-action}
\end{align}
Hence, every even Chevalley generator makes changes in two consecutive rows of the GZ-pattern (with already fairly complicated matrix elements).
The odd Chevalley generators $E_{\bar 1 1}$ and $E_{1 \bar 1}$ are part of the set of generators of Theorem~\ref{theo4}, and they make a change in row~1 only.

Secondly, the highest weight vector of $V([m]^r)$ is given by the pattern in which the labels of the top row are appropriately repeated,
\begin{equation}
v_\Lambda =
 \left|
\begin{array}{lcll:llcll}
 m_{\bar n r} & \cdots & m_{\bar 2 r} & m_{\bar 1 r} & m_{1r} & m_{2r} &\cdots & m_{n-1,r} & m_{nr} \\
\uparrow & \cdots & \uparrow &\uparrow &&&\\
 m_{\bar n r} & \cdots & m_{\bar 2 r} & m_{\bar 1 r} & m_{1r} & m_{2r} &\cdots & m_{n-1,r}  &  \\
  &\ddots &\vdots & \vdots & \vdots &\vdots &\iddots &  \\
&& m_{\bar 2 r} &  m_{\bar 1 r} & m_{1r} & m_{2r}& & &  \\
&&\uparrow & \uparrow \\
&& m_{\bar 2 r} &  m_{\bar 1 r} & m_{1r} && & &  \\
&&&&\downarrow \\
&&& m_{\bar 1 r} & m_{1r} & & & &  \\
&&& \uparrow\\
&&& m_{\bar 1 r }  & & & & & 
\end{array}
\right).
\end{equation}
It is indeed easy to see by means of the previous observation (and the basic actions of Theorem~\ref{theo4}) that 
$E_{\bar 1 1} v_\Lambda = E_{i,i+1} v_\Lambda=0$ ($i=-n,\ldots,-2;1,\ldots,n-1$).
Note that with the diagonal action~\eqref{Ekk}-\eqref{E_nk} one finds
\[
E_{ii} v_\Lambda = m_{ir} v_\Lambda \qquad(i=-n,\ldots,-1;1,\ldots,n).
\]

\setcounter{equation}{0}
\section{The Lie superalgebra $\mathfrak{gl}(\infty|\infty)$} \label{sec:gl(InftyInfty)}%

In this section and in the next one we shall define the Lie superalgebra $\mathfrak{gl}(\infty|\infty)$
and construct a class of $\mathfrak{gl}(\infty|\infty)$ representations.

Consider the set of all squared infinite matrices of the form
\begin{equation}
\left( \begin{array}{ccc:ccc}
 & \vdots & \vdots & \vdots & \vdots & \\
\cdots & a_{\bar 2 \bar 2} & a_{\bar 2 \bar 1} & a_{\bar 2 1} & a_{\bar 2 1} & \cdots \\
\cdots & a_{\bar 1 \bar 2} & a_{\bar 1 \bar 1} & a_{\bar 1 1} & a_{\bar 1 2} & \cdots \\
\hdashline
\cdots & a_{1 \bar 2} & a_{1 \bar 1} & a_{1  1} & a_{1 2} & \cdots \\
\cdots & a_{2 \bar 2} & a_{2 \bar 1} & a_{2  1} & a_{2  2} & \cdots \\
 & \vdots & \vdots & \vdots & \vdots &
\end{array} \right),
\label{inf-mat}
\end{equation}
where the indices take values in the set $\{\ldots, -3,-2,-1;1,2,3,\ldots\}={\mathbb Z}^*\equiv {\mathbb Z}\setminus \{0\}$.
The infinite-dimensional general linear Lie superalgebra 
$\mathfrak{gl}(\infty|\infty)=\mathfrak{gl}_{\bar 0}(\infty|\infty)\oplus \mathfrak{gl}_{\bar 1}(\infty|\infty)$ 
can be defined~\cite{Kac3} as the set of all squared infinite matrices 
of the form~\eqref{inf-mat} such that each matrix has only a finite number of 
nonzero entries, and with grading determined by
\begin{align}
& \mathfrak{gl}_{\bar 0}(\infty|\infty)= \{ (a_{ij})| \hbox{ a finite number of }a_{ij}\in \mathbb{C}\hbox{ are nonzero, and }
a_{ij}=0 \hbox{ when }i\cdot j<0 \};
\label{infgl0}\\
&\mathfrak{gl}_{\bar 1}(\infty|\infty)= \{ (a_{ij})| \hbox{ a finite number of }a_{ij}\in \mathbb{C}\hbox{ are nonzero, and }
a_{ij}=0 \hbox{ when }i\cdot j>0 \}.
\label{infgl1}
\end{align}  
A basis for $\mathfrak{gl}(\infty|\infty)$ is given by all Weyl matrices
$E_{ij}$ ($i,j\in {\mathbb Z}^*$). 
The multiplication  $\llbracket \;,\; \rrbracket$ on $\mathfrak{gl}(\infty|\infty)$ 
is a linear extension of the relations
\begin{equation}
\llbracket E_{ij}, E_{kl}\rrbracket = \delta_{jk}E_{il}-(-1)^{\deg(E_{ij})\deg(E_{kl})} \delta_{il}E_{kj}, \quad 
 i,j,k,l \in  {\mathbb Z}^*.
\label{infEij}
\end{equation}            
A Cartan subalgebra ${\mathfrak H}$ of $\mathfrak{gl}(\infty|\infty)$ is spanned by the elements 
$E_{jj}$ ($j\in {\mathbb Z}^*$) with a natural order $E_{ii} < E_{jj}$ if $i<j$. Denote by 
$\epsilon_i$ ($i\in {\mathbb Z}^*$) the dual basis $\epsilon_i(E_{jj})=\delta_{ij}.$ The correspondence
between root vectors and roots reads
\begin{equation}
E_{ij} \leftrightarrow \epsilon_i - \epsilon_j \quad (i\neq j, \quad i,j\in  {\mathbb Z}^*).
\label{v_roots}
\end{equation}
The set of elements $\{ \epsilon_i - \epsilon_j| i<j\ (i,j\in {\mathbb Z}^*)\}$ (respectively, 
$\{ \epsilon_i - \epsilon_j| i>j\ (i,j\in {\mathbb Z}^*)\}$) are the positive (respectively, negative) 
roots. 
The linear span of all positive root vectors $\{ E_{ij} | i<j\}$ is denoted by $\mathfrak{n}^+$;
the linear span of all negative root vectors $\{ E_{ij} | i<j\}$ by $\mathfrak{n}^-$.
So $\mathfrak{n}^+$ (respectively $\mathfrak{n}^-$) consists of all strictly upper (respectively lower) triangular matrices of the form~\eqref{inf-mat}.

A representation (or module) $V$ of $\mathfrak{gl}(\infty|\infty)$ is said to be a highest weight module if there exists a vector
$v_\Lambda$ (a highest weight vector) such that
\begin{equation}
\mathfrak{n}^+ v_\Lambda = 0,\qquad E_{ii} v_\Lambda = m_i v_\Lambda \qquad(i\in\mathbb{Z}^*).
\label{hw}
\end{equation}
The infinite sequence $[\ldots,m_{-2},m_{-1};m_1,m_2,\ldots]$ is said to be the highest weight of $V$.

It is beyond our purpose to study all irreducible highest weight representations of $\mathfrak{gl}(\infty|\infty)$.
But in the next section we will consider an important subclass: 
these are infinite-dimensional analogs of the covariant representations of $\mathfrak{gl}(n|n)$.
We shall refer to this class as the ``covariant representations of $\mathfrak{gl}(\infty|\infty)$''.

\setcounter{equation}{0}
\section{A class of irreducible representations of  $\mathfrak{gl}(\infty|\infty)$} \label{sec:infty representations} 

In this section, we shall extend the results from Section~\ref{sec:representations} to the case $n \rightarrow \infty$.
Most aspects of this extension are straightforward, but some arguments need more attention.
The basic idea is that the GZ-patterns~\eqref{mn} consisting of $2n$ rows can be extended to GZ-patterns with an infinite number of rows.
However, not all such infinite GZ-patterns are eligible: only the ``stable'' ones are appropriate as basis vectors of an irreducible module.

Let  
\begin{equation}
[m]\equiv [\ldots, m_{-k},\ldots,m_{-2}, m_{- 1}; m_{1},m_{2},\ldots, m_{k},\ldots ], 
\label{M}
\end{equation} 
be a sequence of nonnegative integers such that
\begin{align}
& m_{i}-m_{i+1}\in{\mathbb Z}_{+}  \qquad (i\in \{\ldots,-3,-2;1,2,\ldots \}), \nonumber\\
& m_{-1}\geq \# \{i:m_{i}>0,\;  i \in {\mathbb Z}_{> 0} \}. \label{Icond1}
\end{align}
The last condition is equivalent with
\begin{equation}
{\rm if}\; m_{-1}=p \; {\rm then}\; m_i=0 (\forall\; i>p). \nonumber
\end{equation} 
The sequence~\eqref{M} will be considered as the highest weight of a $\mathfrak{gl}(\infty|\infty)$ representation $V([m])$.
The basis vectors of $V([m])$ will consist of infinite GZ-patterns of the type~\eqref{mn} with the sequence~\eqref{M} as top row.
These GZ-patterns should -- apart from other conditions -- satisfy a stability condition. 
Before giving a precise definition of the infinite stable GZ-patterns, let us give an example.
Consider the pattern
\begin{equation}
\left|
\begin{array}{cccccc:cccccc}
\cdots&9&8&5&3&3&4&3&1&0&0&\cdots\\
 &\vdots&\vdots&\vdots&\vdots&\vdots&\vdots&\vdots&\vdots&\vdots&\vdots& \\
 & &8&5&3&3&4&3&1&0& & \\
 & &8&5&3&3&4&3&1& & & \\
 & & &5&3&3&4&3&1& & & \\
 & & &5&2&2&3&2& & & & \\
 & & & &4&2&4&2& & & & \\
 & & & &3&1&3& & & & & \\
 & & & & &2&3& & & & & \\
 & & & & &1& & & & & & 
\end{array}
\right).
\end{equation}
Such a pattern is called stable with stability index $N=5$ because for all rows with index $i>N$ (recall that we count rows from bottom to top) the entries in each column $k$ ($k\in\mathbb{Z}^*$) are the same.

We are now in a position to state the extension of the main result of Section~\ref{sec:representations}.

\begin{prop}
To each sequence~\eqref{M} (see~\eqref{Icond1}) there corresponds an irreducible $\mathfrak{gl}(\infty|\infty)$
module $V([m])$ with basis $\Upsilon([m])$, consisting of vectors $|m)$ labelled by infinite stable GZ-patterns: $|m)=$ 
\begin{equation}
 \left|
\begin{array}{llcll:llclll}
\cdots & m_{\bar k} &  \cdots & m_{\bar 2} & m_{\bar 1} & m_{1} & m_{2} &\cdots & m_{k-\varphi} &\cdots\\
  &\ddots &\vdots&\vdots &\vdots & \vdots & \vdots &\vdots &\vdots& \iddots & \\
& m_{\bar k,2k-\varphi} & \cdots & m_{\bar 2,2k-\varphi} & m_{\bar 1,2k-\varphi} & m_{1,2k-\varphi} & m_{2,2k-\varphi} &\cdots & m_{k-\varphi,2k-\varphi}    \\
 &  &\ddots &\vdots & \vdots & \vdots &\iddots & & \\
&&& m_{\bar 2,4} &  m_{\bar 1,4} & m_{14} & m_{24}& & &  \\
&&&\uparrow & \uparrow \\
&&& m_{\bar 2,3} &  m_{\bar 1,3} & m_{13} && & & & \\
&&&&&\downarrow \\
&&&& m_{\bar 1,2} & m_{12} & & & & & \\
&&&& \uparrow\\
&&&& m_{\bar 1,1}  & & & & & &
\end{array}
\right)
=
\left| \begin{array}{l} 
[m] \\{}
\vdots \\{}
[m]^{2k-\varphi} \\{}
\vdots \\{}
[m]^{4} \\{}
\\{}
[m]^{3} \\{}
\\{}
 [m]^{2} \\{}
\\{}
[m]^{1}
 \end{array} \right)
\label{IMn}
\end{equation}
satisfying the conditions
\begin{equation}
 \begin{array}{rl}
1. & \hbox{for each table} \; |m) \; \hbox{there exists a nonnegative  (depending on }\; |m) ) \;
\hbox{integer} \; N[|m)], \hbox{called stability} \\
& \hbox{index, such that} \\
&  m_{i,2k-\varphi}=m_{i}, \; \forall\; 2k-\varphi >N[|m)], \; \varphi\in\{0,1\}, \; i\in \{-k,\ldots,-1, 1,\ldots, k-\varphi\};\\
2. & m_{i,2k-\varphi}\in{\mathbb Z}_{+}, \; k\in{\mathbb Z}_{> 0},\; \varphi \in \{0,1\}, \; i\in \{-k,\ldots,-1,1,\ldots, k-\varphi\};\\		
3. & m_{i,2k-\varphi}-m_{i+1,2k-\varphi}\in{\mathbb Z}_{+}, \; k\in{\mathbb Z}_{> 0},\; 
\varphi \in \{0,1\}, \; i\in \{-k,\ldots,-2,1,\ldots, k-\varphi-1\};\\		
4.& m_{i,2k}-m_{i,2k-1}\equiv\theta_{i,2k-1}\in\{0,1\},\quad k\in {\mathbb Z}_{>0}, \; i\in \{-k,\ldots,-1\} \\
5.& m_{i,2k}-m_{i,2k+1}\equiv\theta_{i,2k}\in\{0,1\},\quad k\in {\mathbb Z}_{>0}, \; i\in \{1,2,\ldots,k\}, \\  
6.  &   m_{-1,2k-\varphi}\geq \# \{i:m_{i,2k-\varphi}>0,\; \varphi \in \{0,1\}, \; i\in \{1,2,\ldots,k-\varphi\},\\  
7.& m_{i,2k}-m_{i,2k-1}\in{\mathbb Z}_{+} \;\hbox{ and } \; m_{i,2k-1}-m_{i+1,2k}\in{\mathbb Z}_{+},
\; k\in \{2,3,\ldots\}, \; i\in \{1,2,\ldots, k-1\},\\
8. & m_{i,2k+1}-m_{i+1,2k}\in{\mathbb Z}_{+} \;\hbox{ and } \; m_{i+1,2k}-m_{i+1,2k+1}\in{\mathbb Z}_{+},
\; k\in{\mathbb Z}_{> 0}, \; i\in \{-k-1,-k,\ldots,-2\}.
 \end{array}
\label{ICond3}
\end{equation}
Using the same convention as in~\eqref{lir}, and denoting by $|m)_{\pm (ij)}$ the pattern obtained
from $|m)$ by replacing the entry $m_{ij}$ by $m_{ij}\pm 1$, 
the transformation of the basis under the action of the $\mathfrak{gl}(\infty|\infty)$ generators is as follows ($i\in {\mathbb Z}_{> 0} $)
\begin{align}
& E_{-i,-i}|m)=\left(\sum_{j\neq 0=-i}^{i-1} m_{j,2i-1}-\sum_{j\neq 0=-i+1}^{i-1} m_{j,2i-2}\right)|m); \label{IEKK}\\
& E_{ii}|m)=\left(\sum_{j\neq 0 =-i}^{i} m_{j,2i}-\sum_{j\neq 0=-i}^{i-1} m_{j,2i-1}\right)|m); 
 \label{IENK}
\end{align}
\begin{align}
& E_{-i,i}|m)=\sum_{k=-i}^{-1} \theta_{k,2i-1}
S\Bigl({\sum_{s=1}^{i-1}\sum_{j=-s}^{-1}\theta_{j,2s-1}+\sum_{s=1}^{i-1}\sum_{j=1}^{s}\theta_{j,2s}+\sum_{j=-i}^{k-1}\theta_{j,2i-1}}\Bigr) \nonumber\\
&\times \left(\frac{
{\prod_{j=-i+1}^{-1} (l_{k,2i}-l_{j,2i-2}-1)
\prod_{j=1}^{i} (l_{k,2i}-l_{j,2i})}}
{\prod_{j\neq k=-i}^{-1} (l_{k,2i}-l_{j,2i}) \prod_{j=1}^{i-1}
(l_{k,2i}-l_{j,2i-2}-1) } \right)^{1/2}|m)_{+(k,2i-1)}\; \nonumber\\
& +\sum_{k=1}^{i-1} \theta_{k,2i-2}
S\Bigl({\sum_{s=1}^{i-1}\sum_{j=-s}^{-1}\theta_{j,2s-1}+\sum_{s=1}^{i-2}\sum_{j=1}^{s}\theta_{j,2s}+\sum_{j=1}^{k-1}\theta_{j,2i-2}}\Bigr) \nonumber\\
&\times \left(-\frac{
{\prod_{j=-i+1}^{-1} (l_{j,2i-2}-l_{k,2i-1}+1)
\prod_{j=1}^{i} (l_{j,2i}-l_{k,2i-1})}}
{\prod_{j=-i}^{-1} (l_{j,2i}-l_{k,2i-1})\prod_{j\neq k=1}^{i-1}
(l_{j,2i-2}-l_{k,2i-1}+1) } \right)^{1/2}|m)_{+(k,2i-1)},\; \label{IENI}
\end{align}
\begin{align}
& E_{-i-1,i}|m)=\sum_{k=-i}^{-1} \theta_{k,2i-1}
S\Bigl({\sum_{s=1}^{i-1}\sum_{j=-s}^{-1}\theta_{j,2s-1}+\sum_{s=1}^{i-1}\sum_{j=1}^{s}\theta_{j,2s}+\sum_{j=-i}^{k-1}\theta_{j,2i-1}}\Bigr) \nonumber\\
&\times \left(-\frac{
{\prod_{j=-i-1}^{-1} (l_{k,2i}-l_{j,2i+1})
\prod_{j=1}^{i-1} (l_{k,2i-1}-l_{j,2i-1})}}
{\prod_{j\neq k=-i}^{-1} (l_{k,2i-1}-l_{j,2i-1})\prod_{j=1}^{i}
(l_{k,2i-1}-l_{j,2i+1}+1) } \right)^{1/2}|m)_{-(k,2i)}\; \nonumber\\
& +\sum_{k=1}^{i} \theta_{k,2i}
S\Bigl({\sum_{s=1}^{i}\sum_{j=-s}^{-1}\theta_{j,2s-1}+\sum_{s=1}^{i-1}\sum_{j=1}^{s}\theta_{j,2s}+\sum_{j=1}^{k-1}\theta_{j,2i}}\Bigr) \nonumber\\
&\times \left(\frac{
{\prod_{j=-i-1}^{-1} (l_{j,2i+1}-l_{k,2i+1})
\prod_{j=1}^{i-1} (l_{j,2i-1}-l_{k,2i})}}
{\prod_{j=-i}^{-1} (l_{j,2i-1}-l_{k,2i+1}+1)\prod_{j\neq k=1}^{i}
(l_{j,2i+1}-l_{k,2i+1}) } \right)^{1/2}|m)_{-(k,2i)};\label{IENI1}
\end{align}
\begin{align}
& E_{i,-i}|m)=\sum_{k=-i}^{-1} (1-\theta_{k,2i-1})
S\Bigl({\sum_{s=1}^{i-1}\sum_{j=-s}^{-1}\theta_{j,2s-1}+\sum_{s=1}^{i-1}\sum_{j=1}^{s}\theta_{j,2s}+\sum_{j=-i}^{k-1}\theta_{j,2i-1}}\Bigr) \nonumber\\
&\times \left(\frac{
{\prod_{j=-i+1}^{-1} (l_{k,2i}-l_{j,2i-2}-1)
\prod_{j=1}^{i} (l_{k,2i}-l_{j,2i})}}
{\prod_{j\neq k=-i}^{-1} (l_{k,2i}-l_{j,2i}) \prod_{j=1}^{i-1}
(l_{k,2i}-l_{j,2i-2}-1) } \right)^{1/2}|m)_{-(k,2i-1)}\; \nonumber\\
& +\sum_{k=1}^{i-1} (1-\theta_{k,2i-2})
S\Bigl({\sum_{s=1}^{i-1}\sum_{j=-s}^{-1}\theta_{j,2s-1}+\sum_{s=1}^{i-2}\sum_{j=1}^{s}\theta_{j,2s}+\sum_{j=1}^{k-1}\theta_{j,2i-2}}\Bigr) \nonumber\\
&\times \left(-\frac{
{\prod_{j=-i+1}^{-1} (l_{j,2i-2}-l_{k,2i-1})
\prod_{j=1}^{i} (l_{j,2i}-l_{k,2i-1}-1)}}
{\prod_{j=-i}^{-1} (l_{j,2i}-l_{k,2i-1}-1)\prod_{j\neq k=1}^{i-1}
(l_{j,2i-2}-l_{k,2i-1}) } \right)^{1/2}|m)_{-(k,2i-1)};\label{IFNI}
\end{align}
\begin{align}
& E_{i,-i-1}|m)=\sum_{k=-i}^{-1} (1-\theta_{k,2i-1})
S\Bigl({\sum_{s=1}^{i-1}\sum_{j=-s}^{-1}\theta_{j,2s-1}+\sum_{s=1}^{i-1}\sum_{j=1}^{s}\theta_{j,2s}+\sum_{j=-i}^{k-1}\theta_{j,2i-1}}\Bigr) \nonumber\\
&\times \left(-\frac{
{\prod_{j=-i-1}^{-1} (l_{k,2i}-l_{j,2i+1}+1)
\prod_{j=1}^{i-1} (l_{k,2i-1}-l_{j,2i-1})}}
{\prod_{j\neq k=-i}^{-1} (l_{k,2i-1}-l_{j,2i-1})\prod_{j=1}^{i}
(l_{k,2i-1}-l_{j,2i+1}+1) } \right)^{1/2}|m)_{+(k,2i)}\; \nonumber\\
& +\sum_{k=1}^{i} (1-\theta_{k,2i})
S\Bigl({\sum_{s=1}^{i}\sum_{j=-s}^{-1}\theta_{j,2s-1}+\sum_{s=1}^{i-1}\sum_{j=1}^{s}\theta_{j,2s}+\sum_{j=1}^{k-1}\theta_{j,2i}}\Bigr) \nonumber\\
&\times \left(\frac{
{\prod_{j=-i-1}^{-1} (l_{j,2i+1}-l_{k,2i+1})
\prod_{j=1}^{i-1} (l_{j,2i-1}-l_{k,2i}+1)}}
{\prod_{j=-i}^{-1} (l_{j,2i-1}-l_{k,2i+1}+1)\prod_{j\neq k=1}^{i}
(l_{j,2i+1}-l_{k,2i+1}) } \right)^{1/2}|m)_{+(k,2i)}.\label{IFNI1}
\end{align}
The highest weight vector $|{\hat{m}})$ is the vector~(\ref{IMn}) for which
\begin{equation}
m_{i,2j-\varphi}=m_i,\quad \forall j\in {\mathbb Z}_{> 0},\quad \varphi \in \{ 0,1\},\quad  i\in \{ -j,  \ldots,-1;1,\ldots, j-\varphi \}. 
\label{Ihwv}
\end{equation}
It has stability index $N[|{\hat{m}})]=0$.
\end{prop}

\noindent {\bf Proof.} 
The proof follows ideas developed for the Lie algebra $\mathfrak{gl}(\infty)$~\cite{Palev2}.
Let
\begin{equation}
|m)\equiv 
\left| \begin{array}{l} [m] \\[2mm]
\vdots \\[2mm]
 [m]^{2n} \\[2mm]
\vdots \\[2mm]
 [m]^{2} \\[2mm]
[m]^{1}
 \end{array} \right) \in \Upsilon([m]).
\label{Short}
\end{equation}
Then
\begin{equation}
 [m]^{2n}\equiv [m_{-n},\ldots,m_{-1};m_1,\ldots,m_n],\; n\in {\mathbb Z}_{>0} 
\end{equation}
is called the $2n^{th}$ signature of $|m)$ and 
\begin{equation}
|m)^{\mathop{\rm up}(2n)}\equiv 
\left| \begin{array}{l} [m] \\[2mm]
\vdots \\[2mm]
 [m]^{j} \\[2mm]
\vdots \\[2mm]
[m]^{2n+2} \\[2mm]
[m]^{2n+1}
 \end{array} \right), \quad
|m)^{\mathop{\rm low}(2n)}\equiv 
\left| \begin{array}{l} [m]^{2n} \\[2mm]
\vdots \\[2mm]
 [m]^{i} \\[2mm]
\vdots \\[2mm]
[m]^{2} \\[2mm]
[m]^{1}
 \end{array} \right)
\label{uplow}
\end{equation} 
are said to be the $2n^{th}$ upper and the  $2n^{th}$  lower part of $|m)$, respectively. 
Consider the subalgebra
\begin{equation}
\mathfrak{gl}(n|n)=\{E_{ij}|i,j=-n,\ldots,-1;1,\ldots n\} \subset \mathfrak{gl}(\infty|\infty). 
\end{equation} 
Let $e$ be any $\mathfrak{gl}(n|n)$ generator or polynomial expression of $\mathfrak{gl}(n|n)$ generators.
Then $e|m)$, $|m)\in \Upsilon([m])$ is a linear combination of vectors from $\Upsilon([m])$
with one and the same $2n^{th}$ upper part $|m)^{\mathop{\rm up}(2n)}$.

Denote the set of all vectors~(\ref{IMn}) with one and the same signature $[m]^i, \; i\geq 2n$ by 
\begin{equation}
\Upsilon([m]^i|i\geq 2n) \subset \Upsilon([m]) \label{upvectors}
\end{equation}
and the linear span of $\Upsilon([m]^i|i\geq 2n)$ by 
\begin{equation}
V([m]^i|i\geq 2n) \subset V([m]). \label{upspace}
\end{equation}
Since the subspace $V([m]^i|i\geq 2n)$ is  $\mathfrak{gl}(n|n)$  invariant (see~(\ref{IEKK})-(\ref{IFNI1})),
it will be convenient to define a truncation of vectors $|\mu)$ from $\Upsilon([m]^i|i\geq 2n)$ to their $2n^{th}$ lower part.
So we denote:
\begin{equation}
g(|\mu))=|\mu)^{\mathop{\rm low}(2n)}, \qquad \forall\; |\mu)\in \Upsilon([m]^i|i\geq 2n).
\end{equation}
If 
\begin{equation}
\Upsilon([m]^{2n})=\{ g(|\mu)) \hbox{ such that }|\mu)\in \Upsilon([m]^i|i\geq 2n) \}
\end{equation}
then $g$ is a bijection from $\Upsilon([m]^i|i\geq 2n)$ to $\Upsilon([m]^{2n})$. 
It is straightforward to see that $\Upsilon([m]^{2n})$ is an odd GZ basis of a 
covariant tensor representation of $\mathfrak{gl}(n|n)$  
with highest weight $[m]^{2n}$. Define the action of $\mathfrak{gl}(n|n)$ on 
$|m) \in \Upsilon([m]^{2n})$ by~(\ref{Ekk})-(\ref{Fni1}). Now the linear span
$V([m]^{2n})$ of $\Upsilon([m]^{2n})$ is a covariant tensor representation of $\mathfrak{gl}(n|n)$ with highest weight
$[m]^{2n}$. Comparing the relations~(\ref{IEKK})-(\ref{IFNI1})  with~(\ref{Ekk})-(\ref{Fni1}) 
we have that 
$V([m]^i|i\geq 2n) \subset V([m]) $
is a covariant tensor representation of $\mathfrak{gl}(n|n)$ with signature $[m]^{2n}$ and an
odd GZ basis $\Upsilon([m]^i|i\geq 2n)$ .

Let $E_{ij}$, $E_{kl}$ be any two generators of $\mathfrak{gl}(\infty|\infty)$ and $|m)\in \Upsilon([m])$ be an arbitrary vector. 
For $n\geq \max(|i|,|j|,|k|,|l|)$, consider $E_{ij}$, $E_{kl}$ as generators of $\mathfrak{gl}(n|n) \subset \mathfrak{gl}(\infty|\infty)$.
Then $|m)$ will be a vector from the $\mathfrak{gl}(n|n)$ covariant tensor module 
$V([m]^i|i\geq 2n)\subset V([m])$ and therefore
\begin{equation}
(E_{ij}E_{kl}-(-1)^{\deg(E_{ij})\deg(E_{kl})}E_{kl}E_{ij})|m) = (\delta_{jk}E_{il}-(-1)^{\deg(E_{ij})\deg(E_{kl})} \delta_{il}E_{kj})|m).
\end{equation}
In such a way one concludes that $V([m])$ is a $\mathfrak{gl}(\infty|\infty)$ module.

Next, we wish to prove irreducibility of $V([m])$.
Consider any two vectors $x,y \in V([m])$
\begin{equation}
x=\sum_{i=1}^j \alpha_i |m^i), \quad y=\sum_{i=j+1}^p \alpha_i |m^i), \quad |m^i)\in \Upsilon([m]), \quad \alpha_i\in \mathbb{C}, i\in 1,\ldots p.
\end{equation}
Let 
\begin{equation}
{\tilde{N}}=\max\{N[|m^i)]|i=1,\ldots,p\}.
\end{equation}
Then all vectors $|m^i)$ with $i=1,\ldots,p$ have one and the same $k$ signatures, for $k>\tilde{N}$ and $|m^i)\in V([m]^k|k>\tilde{N})\subset V([m])$.
Therefore $x,y\in V([m]^k|k>\tilde{N})$. The space $V([m]^k|k>\tilde{N})$  is a covariant tensor module of $\mathfrak{gl}(\tilde{N}|\tilde{N})$
and hence there exists 
a polynomial expression $P$ in the generators of $\mathfrak{gl}(\tilde{N}|\tilde{N})$ such that $y=Px$. 
The conclusion is that $V([m])$ is an irreducible 
$\mathfrak{gl}(\infty|\infty)$ module.

Consider the vector $|\hat m )\in \Upsilon([m])$ (see~(\ref{Ihwv})). From~(\ref{IEKK})-(\ref{IENK}) we have
\begin{equation}
E_{ii}|\hat m )=m_i, \quad \forall i\in {\mathbb Z}^*.
\end{equation}
In addition (see the arguments in the proof of Theorem~\ref{theo4})
\begin{equation}
E_{\bar 1 1} |\hat m ) = E_{i,i+1} |\hat m )=0 \hbox{ for all } i=\ldots,-3,-2;1,2,\ldots.
\end{equation}
Following~\eqref{hw}, the irreducible $\mathfrak{gl}(\infty|\infty)$ module $V([m])$ is a highest weight module with highest 
weight vector $|\hat m )$.
\mybox

\vskip 3mm

To summarize the paper, we have developed a new GZ-basis for covariant representations of the Lie superalgebra $\mathfrak{gl}(n|n)$
according to the chain of Lie superalgebras~\eqref{chain-nn}.
The corresponding GZ-patterns have a nice structure with a left and a right triangular array playing an equivalent role.
The set of generators of $\mathfrak{gl}(n|n)$ for which an action on the GZ-basis can be computed consists of odd root vectors only.
The main characteristic of the new GZ-basis is that it can easily be extended to the infinite rank Lie superalgebra $\mathfrak{gl}(\infty|\infty)$, which is not the case for the familiar GZ-basis~\cite{SV2010} of $\mathfrak{gl}(n|n)$.

There is one final property of the class of covariant representations of $\mathfrak{gl}(n|n)$ or $\mathfrak{gl}(\infty|\infty)$ that we have not mentioned yet, and that is unitarity.
Indeed, introduce an inner product $\langle\;,\;\rangle$ in every $\mathfrak{gl}(n|n)$ module $V([m]^r)$ or in every $\mathfrak{gl}(\infty|\infty)$ module $V([m])$, by imposing that the GZ-basis is orthonormal.
Then it follows from the explicit actions \eqref{Ekk}-\eqref{Fni1} or \eqref{IEKK}-\eqref{IFNI1} that these modules are unitary with respect to the star relation 
\[
E_{ij}^* = E_{ji}
\]
for all indices $i$ and $j$.
Unitarity is an important property in a physical context, in particular for the application that we have in mind: the description of
a parastatistics Fock space with an infinite number of parafermions and parabosons.

\renewcommand{\theequation}{A.{\arabic{equation}}}
\setcounter{equation}{0}
\section*{Appendix}
The identity~\eqref{lagrange-id} can easily be proved using standard interpolation theory.
For the notation and results, see any elementary book on interpolation, e.g.~\cite{Whittaker}.

For a function $f(x)$, the interpolation polynomial $p_n(x)$, of degree~$n$ in~$x$, through the distinct points $(x_0,f(x_0))$, $(x_1,f(x_1))$, $\ldots,$ $(x_n,f(x_n))$ is given by
\begin{equation}
p_n(x) = \sum_{i=0}^n f(x_i) L_i(x),
\label{interpol}
\end{equation}
where the $L_i(x)$ are the Lagrange polynomials of degree~$n$:
\[
L_i(x) = \prod_{\myatop{\scriptstyle j=0}{\scriptstyle j\ne i}}^n \frac{(x-x_j)}{(x_i-x_j)}.
\]
When the function $f$ is at least $(n+1)$ times differentiable, the error term (or remainder term) is written as follows:
\begin{equation}
f(x) = p_n(x) + \frac{f^{(n+1)}(\xi)}{(n+1)!} (x-x_0)(x-x_1)\cdots(x-x_n),
\label{remainder}
\end{equation}
where $\xi$ belongs to the interval containing the data points $x_0,\ldots,x_n$.

Now, let $f$ be the function 
\[
f(x) = \prod_{i=0}^n (x-a_i),
\]
which is itself a polynomial of degree $n+1$.
Then $f^{(n+1)}(x)=(n+1)!$, and~\eqref{interpol}-\eqref{remainder} yields:
\begin{equation}
\prod_{i=0}^n (x-a_i) = \sum_{i=0}^n 
\left(\frac{\displaystyle \prod_{j=0}^n (x_i-a_j)}{\displaystyle \prod_{\myatop{\scriptstyle j=0}{\scriptstyle j\ne i}}^n (x_i-x_j)}\right)
\prod_{\myatop{\scriptstyle j=0}{\scriptstyle j\ne i}}^n (x-x_j)
+ (x-x_0)(x-x_1)\cdots(x-x_n).
\label{special-interpol}
\end{equation}
Taking the coefficient of $x^n$ in the left and right hand side gives
\[
-(a_0+a_1+\cdots+a_n) = \sum_{i=0}^n 
\left(\frac{\displaystyle \prod_{j=0}^n (x_i-a_j)}{\displaystyle \prod_{\myatop{\scriptstyle j=0}{\scriptstyle j\ne i}}^n (x_i-x_j)}\right) -
(x_0+x_1+\cdots+x_n),
\]
yielding~\eqref{lagrange-id}.

\section*{Acknowledgments}
The authors were supported by the Joint Research Project ``Representation theory of Lie (super)al\-ge\-bras and generalized quantum statistics'' in the framework of an international collaboration programme between the Research Foundation -- Flanders (FWO) and the Bulgarian Academy of Sciences. NIS was partially supported by Bulgarian NSF grant DFNI T02/6.


\begin{thebibliography}{99}

\bibitem{Green}
Green H S 1953  
A Generalized Method of Field Quantization
{\em Phys.\ Rev.} {\bf 90} 270-73 

\bibitem{Greenberg}
Greenberg O W and  Messiah A M L 1965
Selection Rules for Parafields and the Absence of Para particles in Nature
{\em Phys.\ Rev.\ B} {\bf 138} 1155-67 

\bibitem{Palev1}
Palev T D 1982 
Para-Bose and para-Fermi operators as generators of orthosymplectic Lie superalgebras
{\em J.\ Math.\ Phys.} {\bf 23} 1100-02 

\bibitem{SJ15}
Stoilova N I  and  Van der Jeugt J 2015
Explicit infinite-dimensional representations of the Lie superalgebra $osp(2m + 1|2n)$ and the parastatistics Fock space
{\em J.\ Phys.\ A: Math.\ Theor.} {\bf 48}  155202 (16pp)

\bibitem{parafermion}
Stoilova N I and Van der Jeugt J 2008   
The parafermion Fock space and explicit $\mathfrak{so}(2n+1)$ representations
{\em J.\ Phys.\ A: Math.\ Theor.} {\bf 41} 075202 (13 pp) 

\bibitem{paraboson}
Lievens S, Stoilova N I and Van der Jeugt J 2008  
The paraboson Fock space and unitary irreducible representations of the Lie superalgebra $\mathfrak{osp}(1|2n)$
{\em Commun.\ Math.\ Phys.} {\bf  281} 805-26 

\bibitem{SV2009}
Stoilova N I and  Van der Jeugt J 2009
Parafermions, Parabosons and Representations of $so(\infty)$ and $osp(1|\infty)$
{\em Intern.\ J.\ Math.} {\bf 20} 693-715 

\bibitem{SV2010}
Stoilova N I  and Van der Jeugt J 2010
Gel'fand-Zetlin basis and Clebsch-Gordan coefficients for covariant representations of the Lie superalgebra $gl(m|n)$
{\em J.\ Math.\ Phys.} {\bf 51} 093523 (15 pp)

\bibitem{Kac3}
Kac V G and Van De Leur W 1987  
Super boson-fermion correspondence
{\em Annales de l'institut Fourier} {\bf 37} 99-137 

\bibitem{Kac1}
Kac V G 1977
Lie superalgebras  
{\em Adv.\ Math.} {\bf 26} 8-96 

\bibitem{Kac2}
Kac V G 1978  
Representations of classical Lie superalgebras
{\em Lect.\ Notes Math.} {\bf 676} 597-626  

\bibitem{BR}
Berele R and Regev A 1983 
Hook Young diagrams, combinatorics and representations of Lie superalgebras 
{\em Bull.\ Amer.\ Math.\ Soc.\ (N.S.)} {\bf 8}  No.~2   337-39; \\
Berele R and  Regev A 1987
Hook Young diagrams with applications to combinatorics and to representations of Lie superalgebras 
{\em Adv.\ in Math.} {\bf 64}  No.~2, 118-75 

\bibitem{Macdonald}
Macdonald I G 1995 
{\em Symmetric Functions and Hall Polynomials} 2nd edn (Oxford: Oxford University Press)

\bibitem{JHKR}
Van der Jeugt J, Hughes J W B, King R C and Thierry-Mieg J 1990  
Character formulas for irreducible modules of the Lie superalgebras $sl(m/n)$ 
{\em J.\ Math.\ Phys.} {\bf 18} 2278-2304   

\bibitem{Sergeev}
Sergeev A N 1985
The tensor algebra of the identity representation as a module over the Lie superalgebra $gl(n,m)$ and $Q(n)$  
{\em Math.\ USSR Sbornik} {\bf 51}  419-27 

\bibitem{King}
King R C 1990
$S$-functions and characters of Lie algebras and superalgebras
{\em IMA Vol.\ Math.\ Appl.} {\bf 19}  226-61

\bibitem{Molev} 
Molev A I 2006 
Gel'fand-Tsetlin bases for classical Lie algebras
{\em Handbook of Algebra} {\bf 4} 109-70  
 
\bibitem{Palev2}
Palev T D  1990 
Highest weight irreducible unitary representations of Lie algebras of infinite matrices. I. The algebra $gl(\infty)$
{\em J.\ Math.\ Phys.} {\bf 31} 579-86 

\bibitem{Whittaker}
Whittaker E T and Robinson G 1967
{\em The Calculus of Observations: A Treatise on Numerical Mathematics}
4th edition (New York: Dover)
 
\end{thebibliography}
\end{document}